\def\kms {\hbox{${\rm km\, s}^{-1}$}} % km s-1 for LATEX
\def\cmsq  {$\hbox{{\rm cm}}^{-2}$}    %cm-2
\def\percc {$\hbox{{\rm cm}}^{-3}$}    %cm-3
\def\MOLH {\hbox{${\rm H}_2$}}  %H2
\def\MOLO {\hbox{${\rm O}_2$}}  %O2
\documentclass[12pt,preprint]{aastex}

\shorttitle{Detection of FeO}
\shortauthors{Walmsley et al.}

\begin{document}


\title{Detection of FeO towards SgrB2}


\author{C.M. Walmsley}
\affil{Osservatorio Astrofisico di Arcetri, Largo E. Fermi 5, I-50125
Firenze, Italy
}
\email{walmsley@arcetri.astro.it}

\author{R. Bachiller}
\affil{Observatorio Astron\'omico Nacional (IGN), Campus Universitario,
E-28800 Alcal\'a de Henares (Madrid), Spain}
\email{bachiller@oan.es}

\author{G. Pineau des For\^{e}ts}
\affil{IAS, Universit\'{e} de Paris-Sud, Bat. 121, F-92405, Orsay, France}
\email{forets@mesiob.obspm.fr}

\and

\author{P. Schilke}
\affil{Max Planck Institut f\"{u}r Radioastronomie, Auf dem H\"{u}gel 69,
D-53121 Bonn, Germany} 
\email{schilke@mpifr-bonn.mpg.de}


%\altaffiltext{1}{Visiting Astronomer, Cerro Tololo Inter-American Observatory.
%CTIO is operated by AURA, Inc.\ under contract to the National Science
%\altaffiltext{2}{Society of Fellows, Harvard University.}
%\altaffiltext{3}{present address: Center for Astrophysics,
%    60 Garden Street, Cambridge, MA 02138}
%\altaffiltext{4}{Visiting Programmer, Space Telescope Science Institute}
%\altaffiltext{5}{Patron, Alonso's Bar and Grill}


\begin{abstract}
 We have observed the J=5-4 ground state transition of FeO at a
 frequency of 153 GHz towards  a selection of galactic sources. 
 Towards  the galactic center source
 SgrB2,  we see weak  absorption at approximately the
 velocity of other features towards this source ( 62 \kms \ LSR). 
 Towards other sources, the results were negative as they were also
 for MgOH(3-2) and FeC(6-5). We
 tentatively conclude that the absorption seen toward SgrB2
 is due to FeO in the hot ($\sim $ 500~K)
 relatively
 low density absorbing gas known to be present in this
 line of sight.
 This is the first (albeit tentative) detection of FeO or any iron--containing
 molecule in the interstellar gas. Assuming
 the observed absorption to be due to FeO,
 we estimate [FeO]/[SiO] to be of order 
 or less than 0.002 and
 [FeO]/[\MOLH] of order $3\, 10^{-11}$.  This is compatible
 with our negative results in other sources.
 Our results suggest that the iron
 liberated from grains in the shocks associated with  SgrB2
 remains atomic and is not processed into molecular form. 
\end{abstract}




\section{Introduction}
 The failure to discover iron--bearing molecules in the interstellar
 medium
 is a long standing puzzle.  It is related to the general problem of the
 depletion of refractory elements (see e.g. Jenkins 1989, Weingartner
 and Draine 1999, Walmsley et al. 1999, Walmsley 2000) 
 within both diffuse and dense 
 molecular clouds.  There is good evidence that the degree of depletion
 is correlated with density in diffuse clouds and that it is extremely
 high within molecular clouds.  In fact, the abundance of gas phase
 silicon appears to be six
 orders of magnitude below the solar Si abundance in some circumstances
 (see also Ziurys, Friberg, \& \ Irvine  1989).  Moreover,
 Turner(1991) has put limits
 on the abundance of a variety of molecules containing Na, Si, Mg, Fe,
 and P showing that the case of silicon is not unusual.  The refractory
 elements  thought to be the main constituents of ``silicate grains'' 
 are even more underabundant in the gas phase of dense molecular clouds
 than they are in the diffuse medium sampled by UV observations. 

  Nevertheless, silicon (essentially in the form SiO) is known to be
  present at a low level in  some molecular clouds associated with 
  outflows (see Bachiller and Tafalla 1999, Bachiller et al. 2001,
  Codella et al. 2001) as well as in photon dominated regions (PDRs,
  Schilke et al. 2001).  The general interpretation of this
  is that a small fraction of Si is returned to the gas phase in
  shocks associated with star formation in molecular clouds
  (see e.g. Caselli, Hartquist, \& \ Havnes 1997, Schilke, 
  Pineau des For\^{e}ts \& \ Walmsley 1997).
  One might therefore naively expect iron
  and magnesium to be also liberated in such shocks and
  hence to be present in the molecular gas phase at the same
  low level. With this in mind, we have started an observational
  program searching for Mg/Fe containing species  associated with
  shocks.  

  This article describes a small search carried out with the
  IRAM 30-m telescope  for iron and magnesium containing species.
 In the course of this, we detected evidence
  for the presence of FeO in the molecular clouds seen in absorption
  along the line of sight towards the continuum source 
  in the vicinity of the galactic center SgrB2-M. 
  SiO is well known along this line of sight (Greaves, Ohishi, \& \
  Nyman  1996,
  H\"{u}ttemeister et al. 1995, Peng, Vogel, \& \ Carlstrom 1995)
  and indeed is more  abundant
  in galactic center clouds in general (H\"{u}ttemeister et al. 1998)
  than in molecular clouds in the solar vicinity. The observed large
  line widths suggest that this may be due to shocks (e.g. Flower,
  Pineau des For\^{e}ts \& \ Walmsley 1995) caused by cloud--cloud
 collisions due to shearing motions in the inner Galaxy.
  It is also possible that  the chemistry is affected by the
  hard X--ray sources in the vicinity of SgrB2  which heat
  and ionize the neighbouring molecular clouds (see Martin--Pintado et
  al. 2000). In any case, one expects a correlation between silicon
  and iron and hence it is reasonable to expect
  to find traces of FeO in such regions. 
  In this letter, we  present the evidence that FeO has been detected 
  in the interstellar medium and briefly mention some consequences for the
  chemistry of iron in molecular clouds.

\section{Observations}
 The main body of our observations were carried out at the IRAM 30-m telescope on
 Oct. 14 2001 and confirmatory observations towards SgrB2 were carried
 out on Dec. 1 2001. We observed ( in October) using four receivers simultaneously
 tuned to the frequencies of MgOH(3-2) (88.932268 GHz; the average of
 the doublet frequencies, Ziurys et al. 1996), FeO 5-4 
 ($\Omega =4$, Allen, Ziurys \&  \ Brown  1996, 153135.273 MHz), FeC(6-5) 
 (240862.951 MHz), and MgOH(9-8)(266.728845 GHz). We used
 the wobbler beam switch
 with a throw of 4 arc minutes.   The facility SIS receivers were used
 tuned in single sideband mode with image rejections of
 order 10dB.
 
 As spectrometers, we
 had available at all frequencies filterbanks with 1 MHz
 channel spectral resolution and 256 MHz bandwidth. We also
 used the autocorrelator split into 5 parts. For the 153 GHz
 FeO line which is the main theme of this article, we used a
 resolution of 320 KHz and a bandwidth of 80 MHz.  Pointing
 checks were taken at intervals of roughly 2 hours during the
 observations and showed deviations less than or of order 5 \arcsec .
 The angular resolution of the IRAM 30-m telescope varies from 27\arcsec
 \ at 88 GHz to 16 \arcsec \ at 153 GHz to 10\arcsec at 240 GHz and 
 9\arcsec \ at 266 GHz.  The corresponding beam efficiencies vary
 between 0.78 at 88 GHz to 0.68 at 153 GHz, 0.5 at 240 GHz, and
 0.45 at 266 GHz.  The forward efficiencies vary between 0.95 at
 88 GHz to 0.93 at 153 GHz and 0.9 at 245 GHz. 

 On Dec. 1, we observed (remotely from Bonn)
only SgrB2-M and pointed on that source. We
 observed simultaneously FeO(5-4) and FeO(8-7) (244992 MHz, HPBW 10\arcsec ) in
 wobbler  switch mode with a throw of 4 \arcmin . The
 1 MHz filters were used.

\section{Observational Results}
  We list the sources and positions observed  in 
  table ~\ref{slist} where we give on the scale of T$_{A}^{*}$ the RMS
  noise  in mK.  One can convert to  main-beam brightness temperature
  multiplying by the ratio of forward and beam efficiencies.
  The results were negative with the exception of the
  spectra in FeO towards SgrB2-M and we now discuss this observation. 
  We note that the FeO(5-4) ($\Omega =4$) frequency which we have used
  differs by roughly 6 MHz from that employed in the FeO survey of Merer,
  Walmsley, \& \  Churchwell (1982).

\begin{figure}
%\plotone{f1.eps}
\plotone{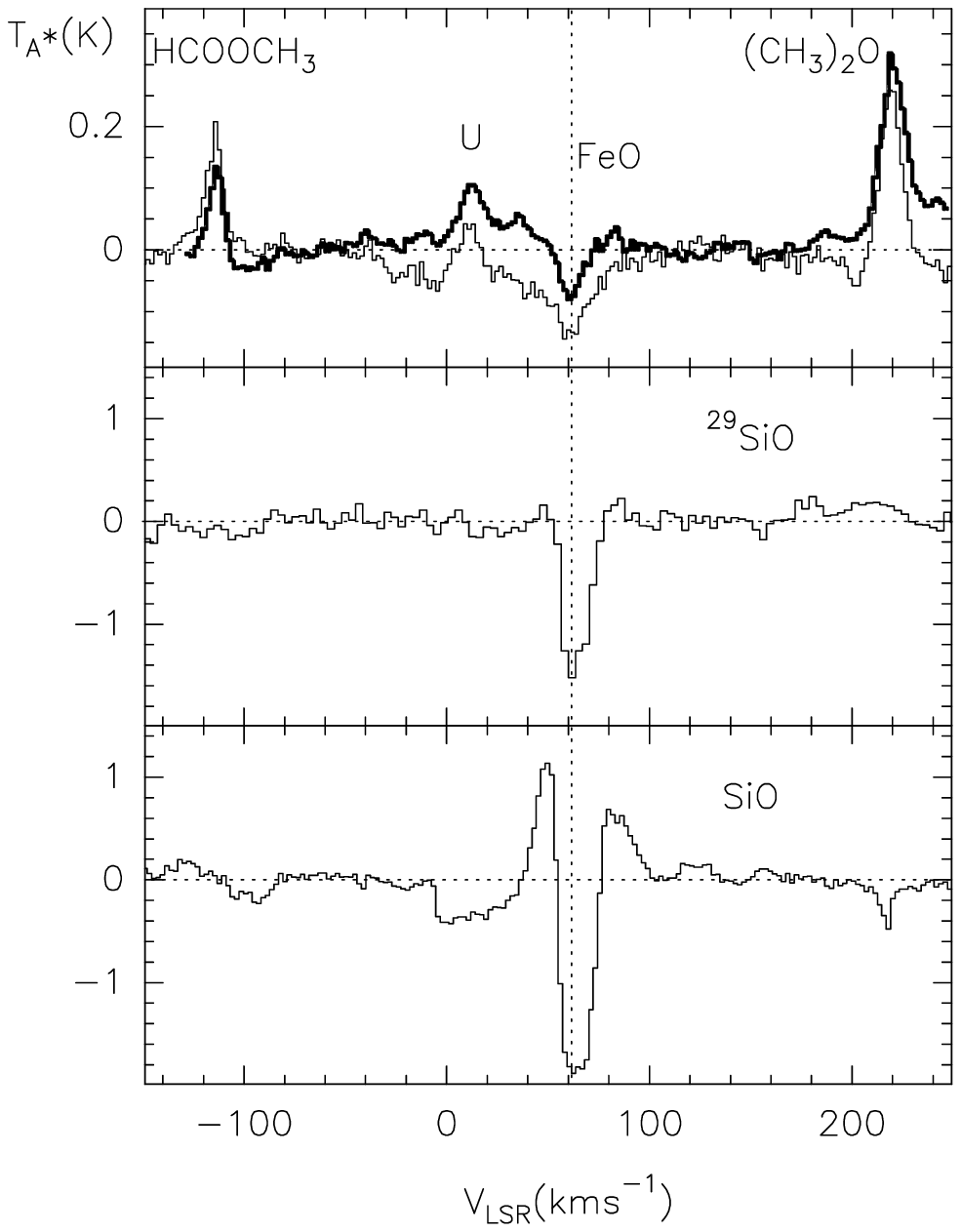}
%\plottwo{revfig.ps}
  \caption
  {The top
panel shows the FeO 5-4 ($\Omega$=4) 
spectrum toward SgrB2--M (1950 coordinates :
R.A. = 17$^{h}$ 44$^{m}$ 10.3$^{s}$,
Dec. = -28$^{\circ}$ 22$^{\prime}$ 04$^{\prime \prime}$,
 full line shows the October spectrum and
bold  line the  Dec.1 measurement). 
%The LSR velocities are with respect to the FeO rest frequency 
%at 153.135273 GHz. 
%Two U-lines are detected
%near 153.226 (probably methyl formate )
%and 153.156 GHz, as well as the 9$_{09}$--8$_{18}$ 
%quadruplet of (CH$_3$)$_2$O (dimethyl ether) at 153.056 GHz.
The middle and lower panels show the 2--1 lines of $^{29}$SiO and
SiO, near 85.759 and 86.846 GHz, respectively, also observed 
with the IRAM 30-m telescope (from de Vicente, 1994).
A dashed line at 61.5 km\,s$^{-1}$ is drawn to illustrate the
good agreement between the main absorption features in the FeO
and  $^{29}$SiO profiles.
    }
  \label{spectrum}
\end{figure}

 Figure ~\ref{spectrum} (top panel)
 shows the spectra taken with the filterbank towards SgrB2-M.
 One sees the 9$_{09}-8_{18}$ quadruplet of
 transitions of dimethyl ether at 153056 MHz as well as a
 line at 153226 MHz (probably HCOOCH$_{3}\, 28_{1,27}-28_{0,28}$)
 and a weak unidentified feature at 153162
 MHz. We superpose the October (thin line,
integration time 36 min. total on plus off)
and December (bold, integr. time 68 min.) spectra for comparison.
 We observe on both dates weak absorption at approximately the
 FeO frequency which stretches over the velocity range
 35-75 
 km s$^{-1}$. The Dec.1 spectrum is reasonably fit with a component of
 width 11.3 \kms \ and velocity 61.3 \kms . 
 The October spectrum also shows traces of a broader
 feature which the Dec.1 data do not confirm and which we presently
 consider to be
 instrumental.  The narrower feature however is present in both
 spectra at the same velocity
 confirming to us that this is an astronomical feature and
 not, for example, of atmospheric origin. 
 The features at 153162 and 153226 MHz are
 also seen in our spectrum towards Orion-KL though not towards
 IRC10216 suggesting to us that they are neither FeO nor carbon--rich
 species.  Neither feature was detected in the FCRAO 14-m survey
 of Ziurys and McGonagle (1993) suggesting that the emission is
 compact. 

 Figure ~\ref{spectrum} (center and lower panels)
 also shows spectra in SiO(2-1) and 
 $^{29}$SiO(2-1) taken by de Vicente (1994) with the IRAM 30-m telescope
 toward SgrB2 with a HPBW of 27\arcsec .  One sees that the velocity of peak
 absorption is at 61.5 \kms \ for both $^{29}$SiO(2-1) and FeO(5-4)
 whereas $^{28}$SiO(2-1) has its peak absorption at higher velocities.
 On the other hand, $^{29}$SiO only shows a narrow (15\kms wide)
 absorption line in contrast  to (optically thick) $^{28}$SiO. 

 We have estimated the area under the ``FeO absorption'' seen in
 Fig. ~\ref{spectrum} as -1.0 K km s$^{-1}$  (in T$_{A}^{*}$ units
 integrating just over the narrow feature with an error of at least 
 50 percent
  mainly dependent on the baseline placement). The continuum
 offset we derive relative to the reference position is
 2.2 $\pm 0.3$~K leading to
 an integrated line-to-continuum ratio of 0.5 km s$^{-1}$ 
 with similar uncertainty. 
 We checked for contamination on the reference position by 
 comparing with a position switched scan with reference 1800\arcsec 
 \ away and found
 no difference within the errors.  We also checked for ``FeO'' 
 emission at positions 20 \arcsec \  offset and found no emission
 greater than 0.15 K.  We conclude therefore that we are observing true
 absorption against the continuum background of SgrB2-M. 
 However, our search for FeO(8-7) towards SgrB2 was negative down to
 a limit of 0.4~K (3~$\sigma $) in T$_{A}^{*}$ units. 
 
 The likelihood
 that we have detected FeO in absorption seems high. In the first place,
 the agreement with the expected line frequency is reasonable.
 There is good agreement with $^{29}$SiO and for $^{28}$SiO
 a difference in the
 peak of the absorption of 4 \kms (v=61 km s$^{-1}$ rather than 
 v=65 km s$^{-1}$ seen in
 SiO(2-1)). While this difference of $\sim $ 2 MHz is well outside
 the uncertainties in the laboratory rest frequencies, 
 the $^{29}$SiO result suggests that it is caused by high
 SiO optical depth (see Peng et al. 1995). 
 %The only other possible ``candidate'' which we have found is the J=4-3
 %transition of AlO which consists of 30 hyperfine components  spread
 %over the frequency range 152987 to 153251 MHz which does not
 %fit particularly well the observed profile. 
% In the second place, there is rough agreement between the
% overall range over which ``FeO absorption'' is seen and the velocity
% range over which SiO(2-1) shows {\it either emission or absorption}. 
% The lower panel of Fig ~\ref{spectrum} shows a spectrum of SiO(2-1) 
% demonstrating the point. Again, it is not clear why the SiO and FeO
% profiles should differ but SiO is highly optically thick
% (see Peng et al. 1995) whereas ``FeO'' is probably not. 
% Moreover, the differences between the October and December profiles
% shows that the ``narrow feature'' of FeO is most reliable. 
Secondly, the FeO(5-4)($\Omega =4$ ) transition observed by us is a
 ground state transition (see Merer et al. 1982) and thus it is
 plausible that one observes absorption towards a
 strong continuum source such as SgrB2-M. Absorption
 towards SgrB2 has been seen in many transitions from the cm range
 (Winnewisser, Churchwell, \& \ Walmsley 1979) to the FIR
 (Ceccarelli et al. 2001)
 and this is attributed generally to the presence of a relatively low
 density ($n(\MOLH )\sim 10^4$ \percc ) hot (500K) foreground layer
 (see e.g. Flower et al. 1995).
 We suspect that we are observing absorption by FeO associated with this
 gas.  The FeO 5-4 transition has a critical density of
 order $10^6$ \percc (Merer et al. 1982 using a dipole moment of
 4.7 Debye from Steimle et al. 1989) and thus it seems reasonable to
 assume that in this foreground layer, FeO is predominantly in the
 J=4 ground state. This is completely consistent with our negative results
 in FeO(8-7).  Using a crude LVG program and supposing collisional
 deexcitation rates of order $10^{-10}$ cm$^{3}$s$^{-1}$,
 we estimate a ratio of roughly 100 between the optical
 depths of the two transitions if the density is $10^4$ \percc .
 This implies an effective optical depth in FeO(8-7) less
 than $10^{-3}$. 

 We note finally that another possible identification might be the AlO(4-3)
 transition which consists of roughly 30 components spread over the
 range 152987 to 153251 MHz. This seems to us unlikely. The pattern does
 not fit well with the observed profile and there is no evidence of an
 analogous feature of AlO(6-5) in the data of Nummelin et al. (1998).

 \section{Analysis}
    Based on the results of the previous section, we now use equation
    2 of Merer et al. to infer the column density of FeO in the
 putative foreground layer. We conclude on this basis that the
 column density $N({\rm FeO})$ (dipole moment of 4.7 Debye from
 Steimle et al. 1989)
 is given by :

 \begin{equation}
  N({\rm FeO}) \, = \, 1.9\, 10^{12} \, \int \tau dv   \ cm^{-2}
  \end{equation}
 where $\tau $ is the optical depth (assumed small) and $v$ is
 the velocity in \kms . We also assume here that the FeO level
 populations are determined by the cosmic 3~K background consistent
 with our negative result in FeO(8-7).  

 We conclude that our results are compatible with a
 FeO column density of order $10^{12}$ \cmsq .  The SiO column density
 on the other hand (Peng et al. 1995 ) is at least
 $5\times 10^{14}$ \cmsq \
 and thus we have [FeO]/[SiO]  of order 0.002.
% This
% conclusion will be reinforced if the absorption observed by us
% is not solely due to FeO. 
 The column density of \MOLH \ in the absorbing layer
 is poorly known  (see Flower et al. 1995)
 but taking a compromise value of $3\times 10^{22}$ \cmsq \ , we
 estimate that [FeO]/[\MOLH] is approximately $3\times 10^{-11}$. 
 
 This is still a
 tiny fraction of the solar iron abundance ($4\times 10^{-5}$ relative
to H) and the obvious question is
 whether FeO is the most abundant gas phase iron component in this hot
 absorbing layer or not.  Little is known about the chemistry of 
 iron--bearing molecules but the frequency sweeps presently available
 (e.g. Nummelin et al.  2000) have not given evidence for molecules
 containing Fe.   Clearly however, searches for species such as
 FeH are warranted and other oxides or hydroxides may be important
 gas phase repositories of iron. We note however that our negative results
 for FeC towards SgrB2 may not be very significant if, as we suspect,
 only the lowest rotational levels are populated in the layer which we
 believe we are detecting in FeO towards SgrB2. 

 Our results towards other sources tell a rather similar story. One may 
 take the case of L1157 as an example where our results put a limit of 
   30 mK (integration time 114 minutes on plus off source,
intensity in main-beam brightness units) on any FeO(5-4) line towards
   the blue--shifted SiO peak (B1 in the nomenclature of Bachiller and 
   P\'{e}rez Guti\'{e}rrez 1997).  For an assumed line width of 10 \kms \ ,
   we derive an upper limit for the FeO column density of $6\, 10^{11}$
   \cmsq \ (assuming an excitation temperature below 50~K). The SiO
   column density in this position is $8\, 10^{13}$ \cmsq \ and so
   we conclude that also towards L1157(B1), one has [FeO]/[SiO] less
   than 0.01. 

  In order to interpret this result, we have constructed a small model
  of iron chemistry in a shock using an updated
  version of the model described by
  Schilke et al.(1997). This assumes that gas phase iron is eroded 
  from grain surfaces due to sputtering in the
  shocked gas (May et al. 2000). We find that while the erosion rates
  are similar for iron and silicon, gas phase iron is much less
  reactive in the shock and in the post shock gas than atomic silicon. 
  This is basically due to the fact that while atomic silicon can react at
  low temperatures with species such as OH and \MOLO \ , the analogous
  reactions for atomic iron (endothermic by 10200~K for Fe+\MOLO \
  and 1550~K for Fe+OH) only occur under high temperature conditions
  in a shock. 
  As a consequence, it seems quite plausible that a few percent of the
  eroded iron atoms are processed into molecular form while essentially
  all of the eroded silicon suffers this fate. We will describe these
  results elsewhere but our provisional conclusion is that the observed
  ratio [FeO]/[SiO] is  explicable in this fashion.  Another interesting
  result is that a considerable fraction of the FeO produced in this
  manner
  can be further processed to FeOH due to the (endothermic) reaction with
  \MOLH . 

\section{Conclusions}

 We believe that we have detected FeO in absorption towards SgrB2. However,
 the feature which we have detected is weak and more confirmatory
 measurements are needed. Searches for other low excitation transitions
 of iron--bearing species would be useful.  A more detailed study of
 the interstellar chemistry of iron is also needed. 

 Even if our identification in SgrB2 turns out to be incorrect, one
 can conclude that in the SgrB2 absorbing layer as well as in the
 post-shock gas which one observes towards L1157, one has
 [FeO]/[SiO] less than 0.01. SiO is thought to be a major form of gas phase
 silicon and thus the SiO abundance gives a measure of silicon depletion.
 For FeO, the preliminary model calculations mentioned earlier 
 suggest that iron is indeed produced by erosion in shocks but
 remains atomic in the post--shock medium.  Indeed [FeII] emission is
 well known in the shocks associated with the Orion outflow
 (Tedds, Brand, \& \ Burton 1999) and so this is quite plausible. We conclude
 therefore that erosion of silicate grains in high velocity (40\kms ) shocks is 
 a plausible explanation of our observation towards SgrB2. 

\acknowledgements 
 We are grateful to the Director of IRAM for granting discretionary
time which enabled us to carry out our observation in Dec. 2001. 
C.M.W acknowledges travel support from the MURST program
"Dust and Molecules in Astrophysical Environments" as well as ASI
grant ARS-98-116.

\clearpage


\clearpage

\begin{table}
\begin{center}
\caption[sources]{Sources and RMS(Noise) for observations of
FeO(5-4), FeC(6-5), and
MgOH(3-2)}
\begin{tabular}{ccccccc}
\tableline
\tableline
  Source  & R.A. (1950) &   Dec. (1950) & Vel.(LSR) & RMS (FeO) $^{a}$ 
  & RMS(FeC) $^{a}$ & RMS(MgOH)$^{a}$\\
	  &  $h \ m \ s$    & $\circ \ \prime \ \prime \prime  $
	  & km s$^{-1}$ &  mK  & mK & mK  \\
\tableline
 W3(OH)   & 02 23 16.5  &   61 38 57    & -45         &  21  & 52 & 17     \\
 Ori-IRc2 & 05 32 47.0  &   -05 24 24   & 7           & 10  & 40 & 7    \\
 IC443-G1 & 06 13 42.0  &  22 33 40     & -10         & 12  & 26 & 7    \\
 IRC10216 & 09 45 14.8  & 13 30 40      & -27         & 9   & 16 & 5       \\
 I16293-E2 & 16 29 27.0 & -24 21 36     & 7           & 8   & 25 & 6  \\
 SgrB2-M   & 17 44 10.3 & -28 22 04      & 65          & 13  & 30 & 17 \\
 SgrB2-N $^{b}$  & 17 44 10.3 & -28 21 14      & 65          &  70 &  & 26  \\
 L1157B   & 20 38 43.1  & 67 50 31      & 1           & 5  & 10 & 3 \\
 \tableline
%\tablenotetext{a}{Noise Values for 1 MHz resolution}
\end{tabular}
\label{slist}
\tablenotetext{a}{Noise Values for 1 MHz resolution}
\tablenotetext{b}{SgrB2-N spectrum is so crowded that the "RMS" given
for FeO is  a measure of confusion due to blended U-lines
rather than noise and FeC(6-5) is
completely blended with U-lines}
\tablecomments{The RMS noise values in columns 5-7 are for
FeO(5-4), FeC(6-5), and MgOH(3-2) respectively}
\end{center}
\end{table}

%% If the table is more than one page long, the width of the table can vary
%% from page to page when the default \tablewidth is used, as below.  The
%% individual table widths for each page will be written to the log file; a
%% maximum tablewidth for the table can be computed from these values.
%% The \tablewidth argument can then be reset and the file reprocessed, so
%% that the table is of uniform width throughout. Try getting the widths
%% from the log file and changing the \tablewidth parameter to see how
%% adjusting this value affects table formatting.

%% In this example, we have used the optional * argument to \\ to
%% instruct LaTeX to keep rows together on the same page. (See the
%% lines following the \cutinhead.) Using \\* to group together table
%% rows on the same page affects how the table breaks. Try taking
%% the *'s out and LaTeXing again to see the difference.

%% Tables may also be prepared as separate files. See the accompanying
%% sample file table.tex for an example of an external table file.
%% To include an external file in your main document, use the \input
%% command. Uncomment the line below to include table.tex in this
%% sample file. (Note that you will need to comment out the \documentclass,
%% \begin{document}, and \end{document} commands from table.tex if you want
%% to include it in this document.)


%% The following command ends your manuscript. LaTeX will ignore any text

\end{document}